\documentclass{ws-rv9x6}
\usepackage{subfigure}   % required only when side-by-side / subfigures are used
\usepackage{ws-rv-thm}   % comment this line when `amsthm / theorem / ntheorem` package is used
\usepackage{ws-rv-van}   % numbered citation & references (default)
\usepackage{braket}
\usepackage{float}
\makeindex
\begin{document}

\chapter[Modelling chemotaxis of microswimmers: from individual to collective behavior]{Modelling chemotaxis of microswimmers: from individual to collective behavior}\label{ra_ch1}

\author[ B. Liebchen and H. L{\"o}wen] { B. Liebchen and H. L{\"o}wen}
%\index[aindx]{Author, F.} % or \aindx{Author, F.}
%\index[aindx]{Author, S.} % or \aindx{Author, S.}

\address{Institut f\"ur Theoretische Physik II: Weiche Materie\\
Heinrich-Heine-Universit\"at D\"usseldorf\\
Universit\"atsstr. 1\\
D - 40225 D\"usseldorf}

\begin{abstract}
We discuss recent progress in the theoretical description of chemotaxis by coupling
the diffusion equation of a chemical species to equations describing the motion of sensing microorganisms.
In particular, we discuss models for autochemotaxis of a single microorganism which 
senses its own secretion leading to phenomena such as self-localization and self-avoidance. 
For two heterogeneous particles, chemotactic coupling can lead to predator-prey behavior including chase and escape phenomena, and to the formation of 
active molecules, where motility spontaneously emerges when the particles approach each other.  
%depending on the sign of the coupling parameter. We then proceed towards a chemotactically
%coupled predator-prey system and discuss conditions leading to trapping and escaping.
We close this review with some remarks on the collective behavior of many particles where 
chemotactic coupling induces patterns involving clusters, spirals or traveling waves. 
\end{abstract}
%\markright{Customized Running Head for Odd Page} % default is Chapter Title.
\body

%\tableofcontents

\section{Introduction}\label{ra_sec1}
Chemotaxis plays a crucial role in the life of many microorganisms. It allows them to navigate towards food sources and away from toxins, but is also used for signaling 
underlying self-organization in multicellular communities. 
Here, microorganisms sense the concentration of a chemical
and adjust their motion to the chemical gradient\cite{Kollmann,Kaupp,Poon}: if the corresponding chemical signal is externally imposed and, say, a food source,
microorganisms will try to move up the gradient towards the food source (``chemoattraction'', or ``positive chemotaxis'').
In the opposite case of a toxin, microorganisms migrate down the gradient (``chemorepulsion'', or ``negative chemotaxis'') \cite{hoell2011theory}. 

Many microorganisms can produce the chemicals to which they respond themselves and use chemotaxis for signaling. Here, 
%Since many microorganisms produce the chemicals to which they respond themselves, 
chemotactic behaviour strongly couples to {\it chemical kinetics}, which is the main topic of the present book.
%The chemical concentration field itself can originate from several sources such that the full
%chemotactic behavior is strongly coupled to {\it chemical kinetics}, the key topic of the present book.

 In this book-chapter we review recent progress in the theoretical description of chemotaxis beyond present
 textbook knowledge. Modelling of chemotaxis concerns the
 coupling between the dynamics of the chemical, as described by the diffusion equation together with appropriate
 source and sink terms, and the motion of microorganisms. Therefore we discuss the basic equations
 for the chemical diffusion and the motion of bacteria (or other microorganisms) which is coupled to the chemical field. 
We then proceed step-by-step from few to many bacteria, where the chemotactic response of a bacterium to chemicals produced by another one
lead to chemical interactions, or signaling, among microorganisms.

 The simplest case of a single bacterium (or "particle") which senses its own secretion is discussed first.
 This case, also called \textit{autochemotaxis},
 is both of biological relevance and of fundamental 
importance as it may lead to effects such as self-localization.
This in turn leads to dynamical scaling laws for the mean-square-displacement of the particle  which are different
 from ordinary diffusion and are therefore of general interest \cite{Metzler2000}. %TODO: No Source??
 We then proceed to a two particle predator and prey system governed by chemotactic sensing
 and finally discuss the general case of
 many (more than two)  particles which probably plays a key role for dynamical cluster formation and other patterns.
Since chemotaxis allows microorganisms to navigate, and also allows to steer synthetic microswimmers, it is 
linked to the 
the rapidly expanding  research field of active particles, 
as for recent reviews see \cite{Elgeti,review_Zoettl,our_RMP,menzel2015tuned}.

\section{Basics: diffusion of chemicals in different spatial dimensions and chemotactic coupling}

\subsection{Diffusing chemicals around static (non-moving) point sources}
To set a theoretical framework for chemotaxis we first explore the kinetics of the chemical that
constitutes the chemotactic signal. We start with the diffusion equation for a chemical
concentration field $c({\vec r}, t)$ in solution
with a point source emitting the chemical with a rate $\lambda_e(t)$, which may generally depend on time, at fixed position ${\vec r}_0$:

\begin{equation} \label{eq:II1}
{{\partial c({\vec r}, t)} \over {\partial t}}
= D_c \Delta c({\vec r}, t) - \mu c({\vec r}, t) + \lambda_e (t)\delta ( {\vec r} -{\vec r}_0)
\end{equation}
\\
Here, $D_c$ is the diffusion coefficient of the chemical in solution.  
The chemical may also evaporate (or disappear) with a rate $\mu$, e.g. due to another chemical reaction.
In the following, we focus on constant emission rates.
The diffusion equation (\ref{eq:II1}) can be considered
in $d=1,2,3$ spatial dimensions which $d=1$ corresponding to an effective slab 
and $d=2$ to an effective cylindrical geometry. Accordingly $\Delta$ denotes the Laplacian 
operator in $d$ spatial dimensions.
For an instantaneous onset of chemical emission at $t=0$, i.e. $\lambda_e (t) = \lambda_e \Theta (t)$,
where $\Theta(t)$ denotes the unit step function and $\mu=0$, the solution of Eqn.~(\ref{eq:II1}) is given in $d$ dimensions by\cite{sengupta2011chemotactic}
\begin{equation} \label{eq:II2}
c(\vec{r},t) = \lambda_e \int^t_0 \text{d}t' \frac{1}{(4\pi D_c |t-t'|)^{\frac{d}{2}}} \exp\left({-\frac{(\vec{r}-\vec{r_0})^2}{4 D_c |t-t'|}}\right) 
\end{equation}
By substituting $t'\rightarrow s:=(\vec{r}-\vec{r}_0)^2/(4 D_c |t-t'|)$, 
this expression can be also written in terms of the upper incomplete Gamma function $\Gamma(a,b)=\int\limits_{b}^{\infty}{\rm e}^{-x} x^{a-1}{\rm d}x $
\begin{equation}
c(\vec{r},t) = \frac{\lambda |\vec{r}-\vec{r}_0|^{2-d}}{4 \pi^{d/2} D_c} \Gamma\left(\frac{d}{2}-2,\frac{(\vec{r}-\vec{r}_0)^2}{4D_c t}\right)
\end{equation}
Expression (\ref{eq:II2}) can be generalized, for $\mu\neq 0$ to 
\begin{equation} \label{eq:II2b}
c(\vec{r},t) = \lambda_e \int^t_0 \text{d}t' \frac{1}{(4\pi D_c |t-t'|)^{\frac{d}{2}}} \exp\left({-\frac{(\vec{r}-\vec{r_0})^2}{4 D_c |t-t'|}-\mu |t-t'|}\right) 
\end{equation}
In many cases, the dynamics of the chemical is fast compared to all other relevant timescales in a given system (e.g. the response time of a microorganism). In these cases, 
we are mainly interested in the chemical steady state profile, corresponding to $\dot c=0$ in Eq.~(\ref{eq:II1}).
This steady state problem is 
%Let us now consider the steady state which is reached after a sufficiently long time after starting the
%secretion assuming a constant $\lambda_e$. This implies that the time derivative on the left handside of eqn. (\ref{eq:II1})
%vanishes.
%Then the problem is 
formally equivalent to screened electrostatics, or in other words, to
linear Debye-H\"uckel theory of screening \cite{HansenL2000} with an inverse 
screening length $\kappa$ now given by

\begin{equation} \label{eq:II2A}
\kappa =\sqrt{\mu/D_c} 
\end{equation} 
\\
while without evaporation ($\mu =0$)
we have an analogy to the Poisson equation of ordinary (unscreened) electrostatics.
Therefore, in various spatial dimensions $d$ the solutions, for a localized initial state, are as follows:
\begin{description}
\item[(i)] In $d=1$, there is an "exponential orbital" around the secreting source fixed 
at the origin of the coordinate system such that for a spatial coordinate $x$ 
the concentration field is for $\mu >0$

\begin{equation} \label{eq:II3}
c(x) =\frac{\lambda_e}{\sqrt{4D_c \mu}} \exp(-\kappa |x|) 
\end{equation}
For $\mu=0$, the steady state solution becomes unphysical; in that case, the time-dependent solution 
does not converge to a steady state but increases forever. 
\\
\item[(ii)] For $d=2$, there is a "Macdonald  orbital"

\begin{equation} \label{eq:II5}
c(r) = {\lambda_e \over D_c} K_0 (\kappa r) 
\end{equation} 
\\
with $r$ denoting the radial distance in two dimensions from the source.
Here $ K_0(x)$ is a Macdonald function (or modified Bessel function)
(see e.g.\ Eqn (39) in \cite{Loewen_JCP_1994}). 
Like in $d=1$, for $\mu=0$ the chemical density does not converge and the steady state solution becomes unphysical.
\\
\item[(iii)] Finally, for $d=3$, there is a radial-symmetric Debye-H\"uckel (or Yukawa) orbital around the point source

\begin{equation} \label{eq:II7}
c(r)   = \frac{\lambda_e}{4 \pi D_c r} \exp (-\kappa r)
\end{equation}
\\
which reduces for $\mu=0$ to the classical Coulomb solution

\begin{equation} \label{eq:II8}
 c(r) = {\lambda_e \over 4\pi D_c r}
\end{equation} 
\\
again with $r$ denoting the radial distance from the point source.
\end{description}

\subsection{Moving point sources}

When the point source is moving with a constant velocity ${\vec v}$, the
 general diffusion equation for constant emission rate is

\begin{equation} \label{eq:II9}
{{\partial c({\vec r}, t)} \over {\partial t}}
= D_c \Delta c({\vec r}, t) - \mu c({\vec r}, t) + \lambda_e \delta ( {\vec r} -{\vec r}_0 - {\vec v}t) 
\end{equation}
\\
By a Galilean transformation from the laboratory frame into the moving particle frame
this equation can be  transformed such that it reads under steady state conditions as follows

\begin{equation} \label{eq:II10}
- D_c \Delta c({\vec r}) + ({\vec v} \cdot \vec{\nabla})  c({\vec r}) + \mu c({\vec r}) =  \lambda_e \delta ( {\vec r})
\end{equation}
\\
Solutions of Eq.~(\ref{eq:II10}) go beyond textbook knowledge and have not been discussed yet in this context. In general, the Green's function
associated with Eq.~(\ref{eq:II10}) can be expressed as a Fourier integral as

\begin{equation} \label{eq:II10A}
c(\vec{r}) = \frac{\lambda}{(2\pi)^d} \int_{-\infty}^{\infty} \text{d}^d k \frac{e^{-i\vec{k}\cdot\vec{r}}}{D_c \vec{k}^2 + i\vec{v}\cdot \vec{k}+\mu}
\end{equation}

Evaluating this integral in one spatial dimension ($d=1$) yields a 
solution consisting of two exponentials with different decay lengths in the front
and in the rear of the moving source:

\begin{equation} \label{eq:II11}
c(x) = \frac{\lambda_e}{\sqrt{4D_c \mu}}
\begin{cases}
     \quad e^{-\chi_+ |x|} \quad\text{for} \quad x\geq 0\\
     \quad e^{-\chi_- |x|} \quad \text{for} \quad x<0
\end{cases} 
\end{equation} 

\begin{equation} \label{eq:II11A}
\text{with} \quad \chi_{\pm} = \sqrt{\frac{\mu}{D_c}+\frac{v^2}{4D_c^2}} \,\pm\, \frac{v}{2D_c}
\end{equation} 

\begin{figure}[]
  \centering
     \includegraphics[width=1.0\textwidth]{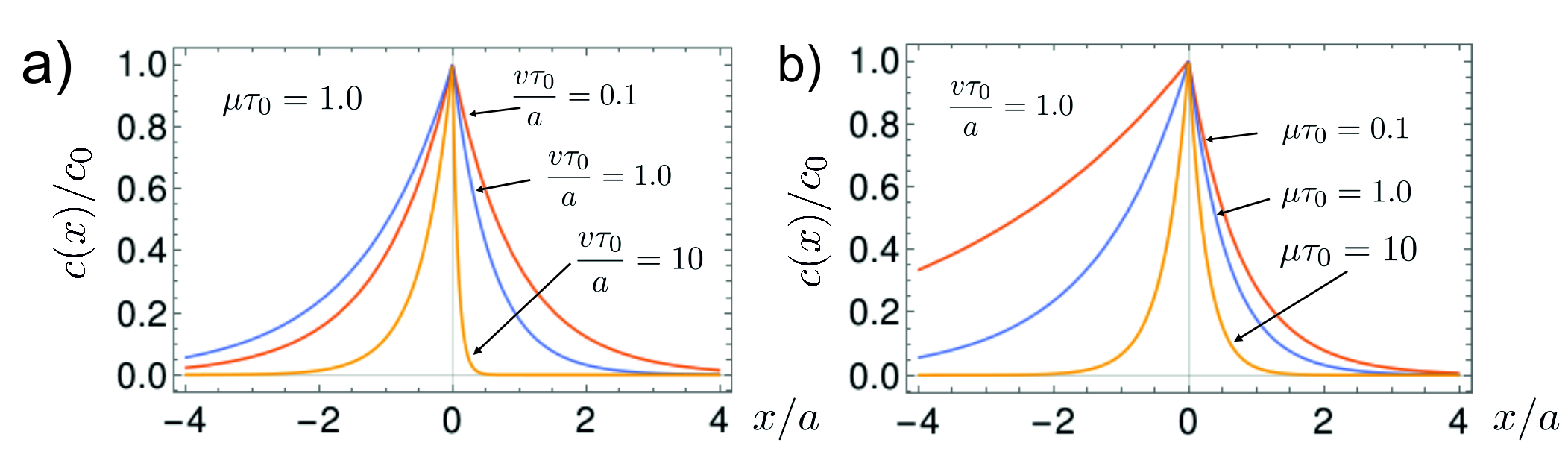}
  \caption{Reduced concentration field $c(x)/c_0$ as a function of reduced distance $x/a$ to the moving source. This is the steady state solution for the one-dimensional diffusion equation of a
 point source moving with a velocity $v$ with constant emission rate $\lambda_e$ for various relative speeds
(a) and relative evaporation rates (b). All length and time scales are given in terms of $a=\sqrt{D_c/\lambda_e}$
and $\tau_0 = 1/\lambda_e$, and $c_0=1/(2a\sqrt{\mu \tau_0})$. 
%$c_0 = 1/(\mu a \tau_0)$. 
The parameters are: (a) $v \tau_0 /a = 0.1, 1.0, 10$ at $\mu \tau_0 = 1$ and
(b) $\mu \tau_0 = 0.1, 1.0, 10$ at $v \tau_0 /a =1$.}
  \label{Fig1}
\end{figure}

This solution is plotted in (\ref{Fig1}). Clearly, increasing the speed of the source enhances the front-rear asymmetry while
at $v=0$ we recover the solution (\ref{eq:II3}) of a static point source. 
Increasing the evaporation rate basically decreases the range of the chemical concentration around the source.

%In two and three dimensions, an analytical solution is not available to the best of our
%knowledge. 
For the corresponding solution in two dimensions, we find: 
\begin{equation}
c(\vec{r})= \frac{\lambda_e}{D_c} K_0(\tilde \kappa r)\exp\left(-\frac{\vec{v}\cdot\vec{r}}{2D_c}\right);\;\; \text{with}\;\; \tilde \kappa=\sqrt{\frac{\mu}{D_c} + \frac{\vec{v}^2}{4D_c^2}}
\end{equation}
and in three dimensions: 
\begin{equation}
c(\vec{r})= \frac{\lambda_e}{4\pi D_c r} \exp\left(-\tilde \kappa r - \frac{\vec{v}\cdot\vec{r}}{2D_c}\right)
\end{equation}
For $\mu = 0$, in a coordinate system whose $x$-axis points along $\vec{v}$, the three dimensional solution reduces to %\cite{sengupta2011chemotactic}
\begin{equation} \label{eq:II13}
c(\vec{r}) = \frac{\lambda_e}{4\pi D_{c}r}\exp\left[-\frac{|\vec{v}|(x+|\vec{r}|)}{2D_{c}}\right]
\end{equation}
%{\bf Different to Sengupta et al. paper -- Here, we have a factor of $4\pi$ instaed of $2\pi$}
%\cite{sengupta2011chemotactic} sengupta2
\\
Remarkably in the rear of the moving source ($x<0$ at $y=z=0$ such that $x+r=0$)
the chemical concentration decays algebraically as $1/|x|$ while in all other directions it decays
with a Yukawa-behavior as $\exp{(-vx/D_c)}/x$, i.e. algebraically for $r\ll \sqrt{D_c/\mu}$ and exponentially at longer distances. 
This asymmetry indicates the significance of the source trail
and a memory effect about the past of the secreting particle.

\subsection{Chemotactic coupling and secreting particle dynamics}

For chemotaxis in its simplest form, the particle directs its motion
according to the gradient of the chemical field. We describe this
coupling to the chemical concentration as an effective force

\begin{equation} \label{eq:II14}
{\vec F} = \alpha \nabla c({\vec r}, t)
\end{equation} 
\\
acting on the particle. Here positive $\alpha$ values represent ``positive'' chemotaxis or ``chemoattraction''whereas 
negative $\alpha$ values represent ``negative'' chemotaxis or ``chemorepulsion''.
The linear coupling to the gradient is the simplest possible form, but
other couplings like logarithmic ones are also conceivable and probably relevant for microbiological systems \cite{MurrayII} where
\begin{equation} \label{eq:II15}
{\vec F} = \alpha \vec{\nabla} \ln c({\vec r}, t) = \alpha   \frac{\vec{\nabla} c({\vec r}, t)} {c({\vec r}, t)}
\end{equation}
A more complicated coupling involves a concentration dependent prefactor $\alpha$
 as  proposed in Ref. \cite{Lutz}. We will basically use and discuss formula (\ref{eq:II14}) in the sequel.

The effective chemotactic force typically acts on a completely overdamped
particle with position ${\vec r}_p (t)$, leading to the following equation of motion: 
\begin{equation} \label{eq:II16}
 \gamma {d \over  dt} {\vec r}_p (t)  = \alpha \nabla c({\vec r}_p(t), t) 
\end{equation} 
\\
Here, $\gamma$ is the Stokes drag coefficient.
If necessary, additional noise terms can be added to Eq.~(\ref{eq:II16})
 in order to model the stochastic collisions
 of the particle with the solvent molecules. Here we neglect any hydrodynamic flow effects stemming from a finite radius of the point
 source are neglected. Their inclusion would require a more sophisticated analysis.

\section{Autochemotaxis for a single particle}
If a single particle emits a chemical to which it responds itself, we call this ``autochemotaxis''.
Here, Tsori and de Gennes \cite{Tsori} have coupled the chemical diffusion equation to an equation of motion for a particle. %TODO
For the chemoattractant case at vanishing evaporation rate  $\mu=0$, they have found "{\it self-trapping}" of
the particle in a spatial dimension $d=1,2$
 but not for $d=3$. This implies that a particle traps itself if moves towards the chemical which it has secreted in the past. 
The concept of ``perfect'' self-trapping was subsequently questioned  by Grima
\cite{Grima1,Grima2} in a model with a positive evaporation rate  $\mu>0$. %TODO
Here it turned out that self-trapping is a transient phenomenon at $\mu>0$ and crosses over to normal diffusion at very
long times even for $d=1,2$. For sufficiently strong negative chemotaxis, Grima found
long-time diffusive or {\it ballistic\/} motion depending on the secretion rate  $\lambda_e$. This result
which was obtained for any dimensionality $d$ suggests that a particle might {\it self-propel\/}
if it avoids the region where it has been in the past, 
and in some sense constitutes a link between repulsive autochemotaxis
and the rapidly growing research field of active particle
or microswimmers \cite{review_Zoettl,menzel2015tuned,Gompper2016}.
(Note however, that while self-propulsion due to autochemorepulsion might apply to particles on a surface, for microorganisms in bulk, which oblige momentum 
conservation, self-propulsion based on autochemorepulsion is conceptually not immediate and would probably need to involve some combination of parity-symmetry breaking 
and phoresis.) 

Now we shall mainly review subsequent studies which include Brownian noise due to solvent molecules acting 
on the self-driven particles. Noise statistics is a relevant part of the actual
trajectories of self-propelled particles \cite{our_RMP} both for microorganisms and synthetic microswimmers
\cite{tenhagen2011brownian,zheng2013non,KuemmeltHWBEVLB2013}. It is expected that noise will destroy the perfect localization for
positive autochemotaxis as well as the ballistic long-time motion for negative autochemotaxis. Indeed
this was confirmed by numerical work and theoretical analysis in a subsequent paper of Sengupta et al
\cite{sengupta2009dynamics} which we shall discuss in the following in more detail.

The governing equations of the Brownian noise model introduced by Sengupta et al  \cite{sengupta2009dynamics} describe
a coupling between the diffusion equation of the chemical concentration field $c({\vec r}, t)$ and the
trajectory of the secreting particle ${\vec r}_p(t)$. At vanishing evaporation rate $\mu$, the
chemical is emitted with a constant rate $\lambda_e$ and is diffusing in $d$ spatial dimensions according to

\begin{equation} \label{eq:III1}
\frac{\partial c(\vec{r},t)}{\partial t} = D_c \nabla^2 c(\vec{r},t)+\lambda_e \delta (\vec{r}-\vec{r_p}(t))
\end{equation} 
\\
%hier Gleichung (1) von \cite{sengupta2009dynamics} einf\"ugen mit der Notations{\"a}nderung $\rho$ -> c,
%boldface f\"ur Vektoren ersetzen durch \vec,
The equation of motion for the emitting particle is given by

\begin{equation} \label{eq:III2}
\gamma \dot{\vec{r}}_p(t) = \vec{F}(\vec{r}_p,t) + \vec{\eta}(t)
\end{equation} 
\\
%TODO
%hier Gleichung (2) von \cite{sengupta2009dynamics} einf{\"u}gen, boldface f\"ur Vektoren ersetzen durch \vec

Here, ${\vec \eta}(t)$ is an effective Gaussian white noise which zero mean and variance

\begin{figure}[]
  \centering
     \includegraphics[width=1.0\textwidth]{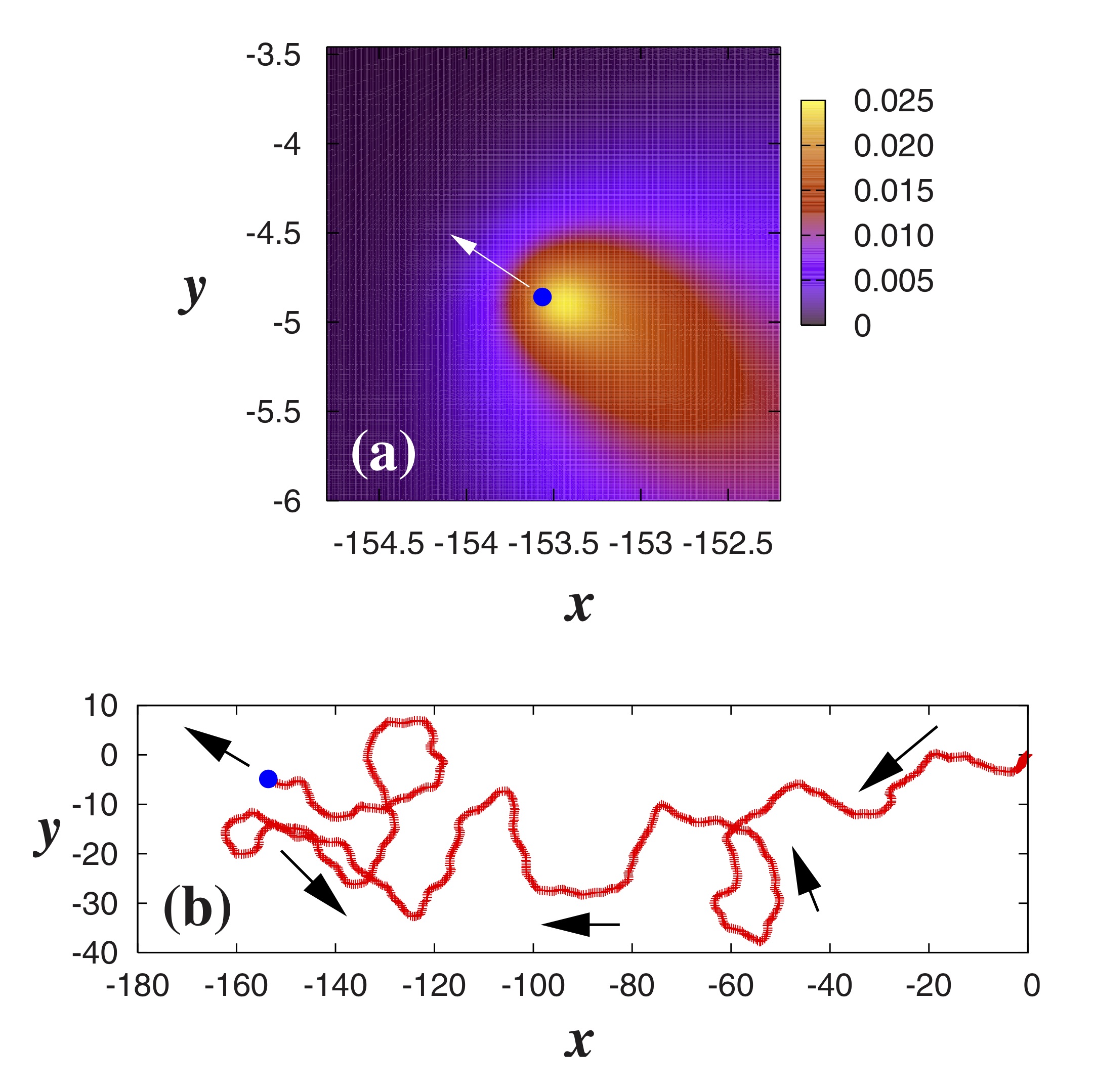}
  \caption{(a) Snapshot of the instantaneous density
profile $c(x,y)$ of the chemorepellent released by the microorganism
moving in two dimensions, obtained from simulation, at time instant $t=10.0$ (in units of $\lambda_e^{-1}$). (b) The entire trajectory, shown as the red (thick) curve, of the microorganism. The current position of the microorganism $\vec{r}_p$  is indicated by the blue (black) dot in both the figures, and the direction of motion is indicated by arrows along the trajectory. The corresponding coupling strength being $|\lambda |=10000$. The parameters are $D \equiv 1/\beta\gamma=0.1 \ell_o^2/\tau_0$, $D_c/D = 100$, $t_0=0.001 \tau_0$ with length, time and energy scales of $\ell_0 = \sqrt { \sqrt{D_c D}/\lambda_e }$, $\tau_0 = 1/\lambda_e$, $1/\beta$. 
Figure from Ref. \cite{sengupta2009dynamics}.}
  \label{Fig2}
\end{figure}

\begin{equation} \label{eq:III2a}
\braket{\eta_i(t)\eta_j(t')} = 2 \gamma \beta^{-1} \delta_{ij}(t-t')
\end{equation} 
\\
%hier Gleichung (2a) von \cite{sengupta2009dynamics} einf\"ugen
%TODO
 with $i$ and $j$ denoting the Cartesian spatial components and $\beta$ an effective inverse thermal energy. In Ref. \cite{sengupta2009dynamics}, the chemotactic force
${\vec F}({\vec r}_p,t)$ depends on the history of the particle trajectory ${\vec r}_p(t)$
apart from a delay (or memory) time $t_0$ which takes into account that a finite time is
needed for sensing the chemical. Integrating or superimposing over the Green's function
of chemical diffusion, Eq.(\ref{eq:II2}), the chemotactic force ${\vec F}({\vec r}_p,t)$, Eq.~(\ref{eq:II14}), is modelled as

\begin{equation} \label{eq:III3}
\vec{F}(\vec{r},t) = -2 \alpha \lambda_e \int_0^{t-t_0} \text{d}t' \frac{(\vec{r}-\vec{r}_p(t'))}{4D_c|t-t'|} \frac{\exp{\left[\frac{-(\vec{r}-\vec{r}_p(t'))^2}{(4D_c|t-t'|)} \right]}}{(4 \pi D_c|t-t'|)^{d/2}}
\end{equation} 
%hier Gleichung (3) von \cite{sengupta2009dynamics} einfügen, boldface für Vektoren ersetzen durch \vec
%TODO

For $d=2$, a typical particle trajectory in the chemorepellent case and the associated chemical density field
are shown in Figure \ref{Fig2}. Clearly the particle avoids its own trail where it had been in the past
giving rise to a persistent random walk along the arrow shown in \ref{Fig2}.

\begin{figure}[]
  \centering
     \includegraphics[width=1.0\textwidth]{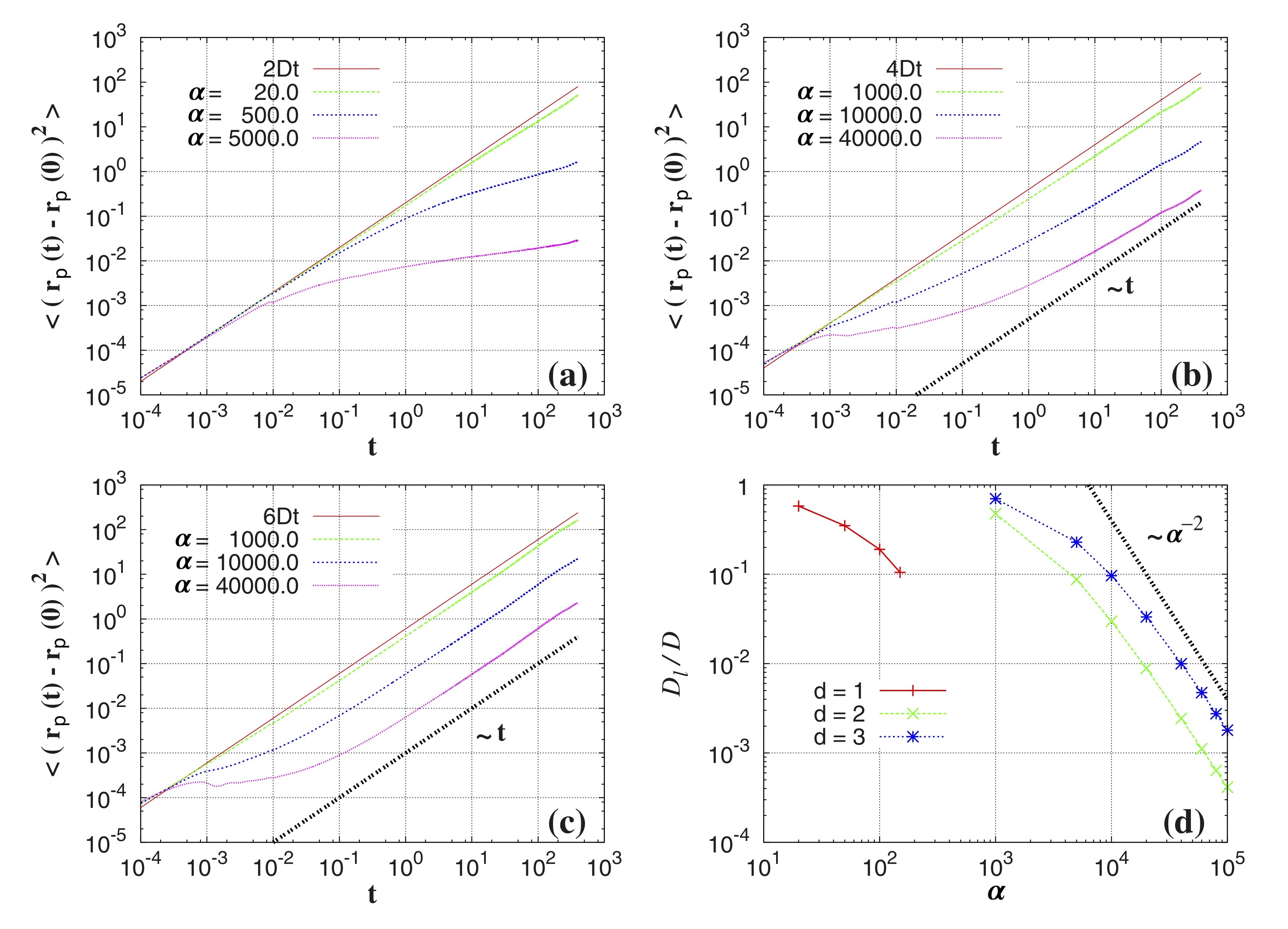}
  \caption{Mean-square displacement $\braket{[\vec{r}_b(t)-\vec{r}_b(0)]^2}$ of the microorganism as a function of time $t$ with chemoattractant in (a) $d=1$ with $\alpha=20,500,5000$; (b) $d=2$ with $\alpha=1000,10000,40000$; (c) $d=3$ with $\alpha =1000,1000,40000$. The nonchemotactic diffusion reference lines are also indicated as $2Dt, 4Dt$, and $6Dt$ correspondingly for $d=1,2,3$. Reference lines (thick dotted) are used to indicate the long-time diffusive behavior (~t) wherever possible. The relative long-time diffusivity $D_l/D$ is shown as a function of $\alpha$ in (d) for $d=1,2,3$. The reference line (thick dotted) shows a power-law scaling behavior $~1/\alpha^2$ (see text). The parameters are as in \ref{Fig2}, $D = 1/\beta\gamma$ is the short-time particle diffusivity and the coupling parameter $\alpha$ is measured in terms of $\ell_0^d / \beta$. 
Figure from Ref\cite{sengupta2009dynamics}. }
  \label{Fig3}
\end{figure}

%Figure 2: hier Figure 5 von \cite{Sengupta1} einfügen mit Caption, in der Caption müssen die Einheiten diskutiert werden
Results for the "exact" numerical solution of these governing equations are presented in Figure \ref{Fig3} for the
chemoattractive case. There is long-time diffusive behaviour with a long-time diffusion coefficient $D_l$
but for stronger couplings $\alpha$, an intermediate transient time region shows up where the particle
is quasi-localized.
This localization is most pronounced in low spatial dimensions $d$.
The long-time self-diffusion coefficient $D_l$ drops strongly with the coupling $\alpha$ for any $d$
(see Figure \ref{Fig3}(d)) and scales with $1/\alpha^2$ in agreement with scaling arguments proposed in
Ref. \cite{sengupta2009dynamics}.

%Figure 3: hier Figure 1 von \cite{Sengupta1} einfügen mit Caption plus:
%The parameters are as in \ref{Fig2}, $D = 1/\beta\gamma$ is the short-time particle diffusivity
 %and the coupling parameter $\lambda$ is measured in terms of $\ell_0^d / \beta$.
%From Ref.\ \cite{sengupta2009dynamics}.

\begin{figure}[H]
  \centering
     \includegraphics[width=1.0\textwidth]{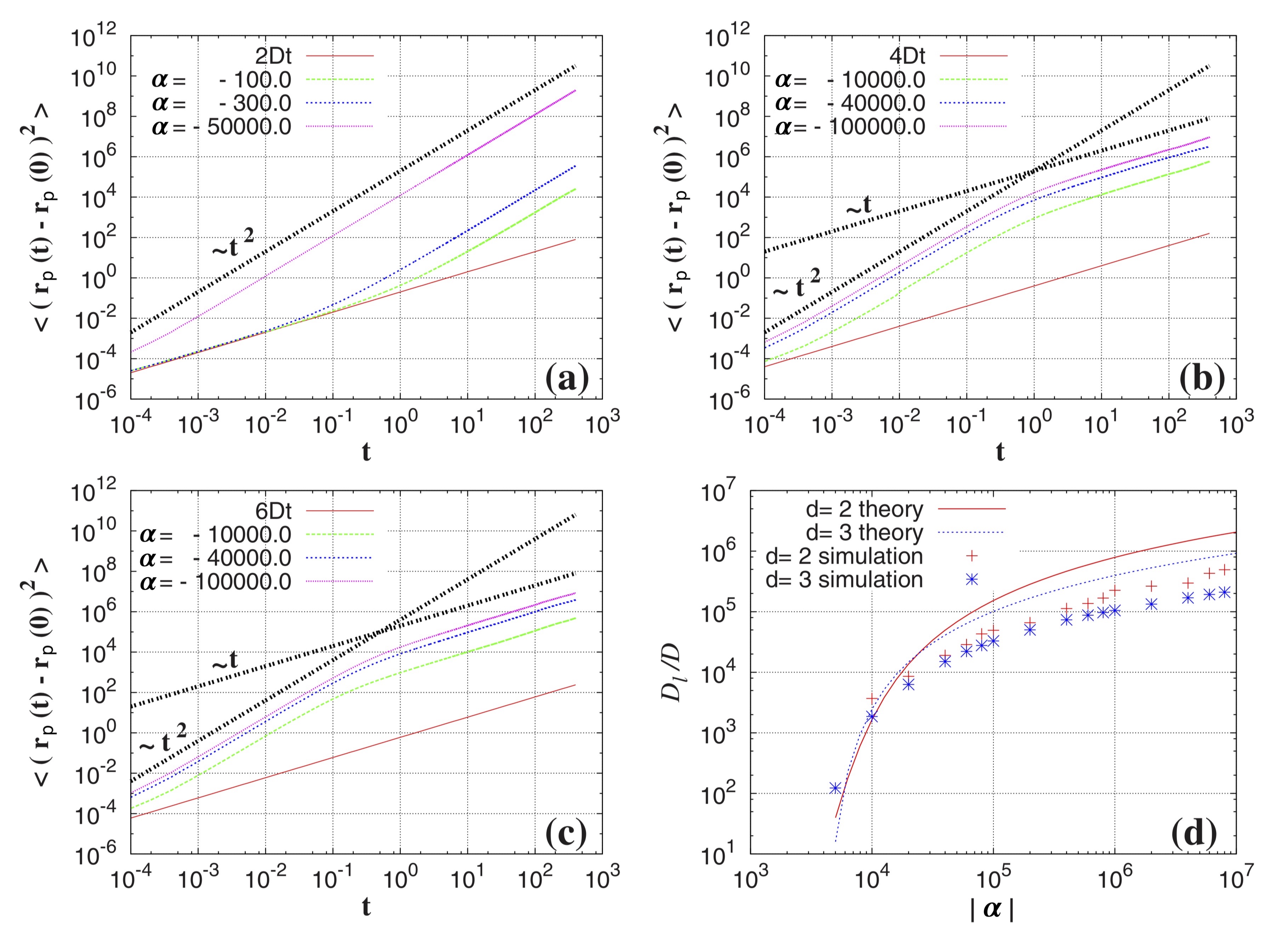}
  \caption{Mean-square displacement $\braket{[\vec{r}_b(t)-\vec{r}_b(0)]^2}$ of the microorganism as a function of time $t$ with chemorepellent in (a) $d=1$ with $\alpha=-100,-300,-50000$; (b) $d=2$ with $\alpha=-1000,-40000,-100000$; (c) $d=3$ with $\alpha =-1000,-40000,-100000$. The nonchemotactic diffusion reference lines are also indicated as $2Dt, 4Dt$, and $6Dt$ correspondingly for $d=1,2,3$. Reference lines (thick dotted) indicating the ballistic (~$t^2$) and the long-time diffusive (~t) dynamics is shown as guide to the eye. The relative long-time diffusivity $D_l/D$ is shown as a function of $|\alpha|$ in (d) for $d=2,3$. The points represent the actual data obtained from simulations, the lines correspond to a semiquantitative theory (see text). The parameters are as in Figure \ref{Fig2}, $D = 1/\beta\gamma$ is the short-time particle diffusivity and the coupling parameter $\alpha$ is measured in terms of $\ell_0^d / \beta$. Figure 
from Ref \cite{sengupta2009dynamics}. }
  \label{Fig4}
\end{figure}

Figure \ref{Fig4} shows the chemorepulsive case. Here, we again have a transient ballistic regime is transient which is most
pronounced in low spatial dimensions. A simple theory put forward in Ref.\cite{sengupta2009dynamics}
describes the strong increase of the long-time particle diffusivity with the coupling strength $|\lambda|$.

%Figure 4: hier Figure 2 von \cite{Sengupta1} einfügen mit Caption plus:
%The parameters are as in Figure 2, $D = 1/\beta\gamma$ is the short-time particle diffusivity
% and the coupling parameter $\lambda$ is measured in terms of $\ell_0^d / \beta$.
%From Ref.\ \cite{Sengupta1}.
At this stage we mention that is is now possible to create synthetic particles which react
in principle such that they avoid their own secretion trail. One important example discussed
recently is an oil droplet in an aqueous
surfactant solution which ``remembers'' the  surfactant concentration which constitutes its self-propagation
\cite{Maass}. Another idea is to dynamically control the motion of colloidal particles by optical fields
which are dynamically adapted (programmed) such that the colloids avoid positions where they have been at earlier times; this leads to 
self-propulsion \cite{Egelhaaf}.
A third realization in the macroscopic world are robots which can be programmed at wish \cite{PRX_2016_Volpe}.
Moreover there are further but related theoretical models for particles avoiding their own past
trails \cite{Kranz,Golestanian2} or  involve memory effects  leading to similar phenomena \cite{Taktikos,chucker}.

\section{Chemotactic predator-prey dynamics} %IV

Next we shall explore two particles which are sensing each other via chemotaxis mimicking signaling among microorganisms.
One particle ("predator") is attracted
by the chemical secreted by the second particle and the latter ("prey")
is repelled by the chemical secreted by the first one. Here we follow the model of Ref.
 \cite{sengupta2011chemotactic}. Now we have two chemicals
characterized by concentration fields
$c_i({\vec r}, t)$ $(i=1,2)$ and two trajectories ${\vec r}_i(t)$.
In the absence of chemical evaporation ($\mu=0$), the concentration fields read

\begin{equation} \label{eq:IV1}
c_i(\vec{r},t) = \lambda_i \int^t_0 \text{d}t' \frac{1}{(4\pi D_{ci} |t-t'|)^{\frac{d}{2}}} \exp\left({-\frac{[\vec{r}-\vec{r_i}(t')]^2}{4 D_{ci} |t-t'|}}\right) 
\end{equation} 
\\
%Hier Gleichung (5) von Ref. \cite{Sengupta2} einfügen (IV.1)
which is the solution of the chemical diffusion equation for given trajectories $\vec{r}_i(t)$ 
with $D_{ci}$ denoting the diffusion coefficient of the two chemicals and $\lambda_i$ being the production rate of chemical species $i$.
The equations of motion determining the predator trajectory reads: 

\begin{equation} \label{eq:IV2}
\gamma_1 \dot{\vec{r_1}} = +\alpha_1 \nabla c_2(\vec{r}_1,t) + \vec{\eta}_1(t)
\end{equation} 
%Hier Gleichung (1) on Ref. \cite{Sengupta2} einfügen (IV.2)
and the equation of motion for the prey is

\begin{equation} \label{eq:IV3}
\gamma_2 \dot{\vec{r_2}} = -\alpha_2 \nabla c_1(\vec{r}_2,t) + \vec{\eta}_2(t)
\end{equation} 
where $\gamma_{1,2}$ are friction coefficients and $\alpha_{1,2}$ are chemotactic coupling coefficients. 

%Hier Gleichung (2) on Ref. \cite{Sengupta2} einfügen (IV.3)
As in Eq.~(\ref{eq:III2}) we generally allow for Gaussian white noise, represented by $\vec{\eta}_{1,2}(t)$.

It is important to remark here that the chemically mediated interaction between the predator and the prey is nonreciprocal, i.e. the force
exerted by the predator acting on the prey is unequal to the force acting on the predator due to the prey particle.
%on the first particle by the presence of the second particle is {\bf not} the
%negative of the force exerted on the second particle by the first particle. 
This violation of
Newton's third law stems from the non-equilibrium conditions and applies to the \emph{effective interaction} 
among predator and prey; the microscopic interactions among all solvent molecules, and the predator and the prey particle 
are of course reciprocal so that momentum conservation applies and any net motion of the predator-prey-pair (in bulk) will be generally 
balanced by a counter-propagating flow of solvent (and or chemicals). Such 
nonreciprocal interactions are frequently encountered
in situations away from equilibrium, e.g. in dusty plasmas \cite{Ivlev}.

\begin{figure}[]
  \centering
     \includegraphics[width=1.0\textwidth]{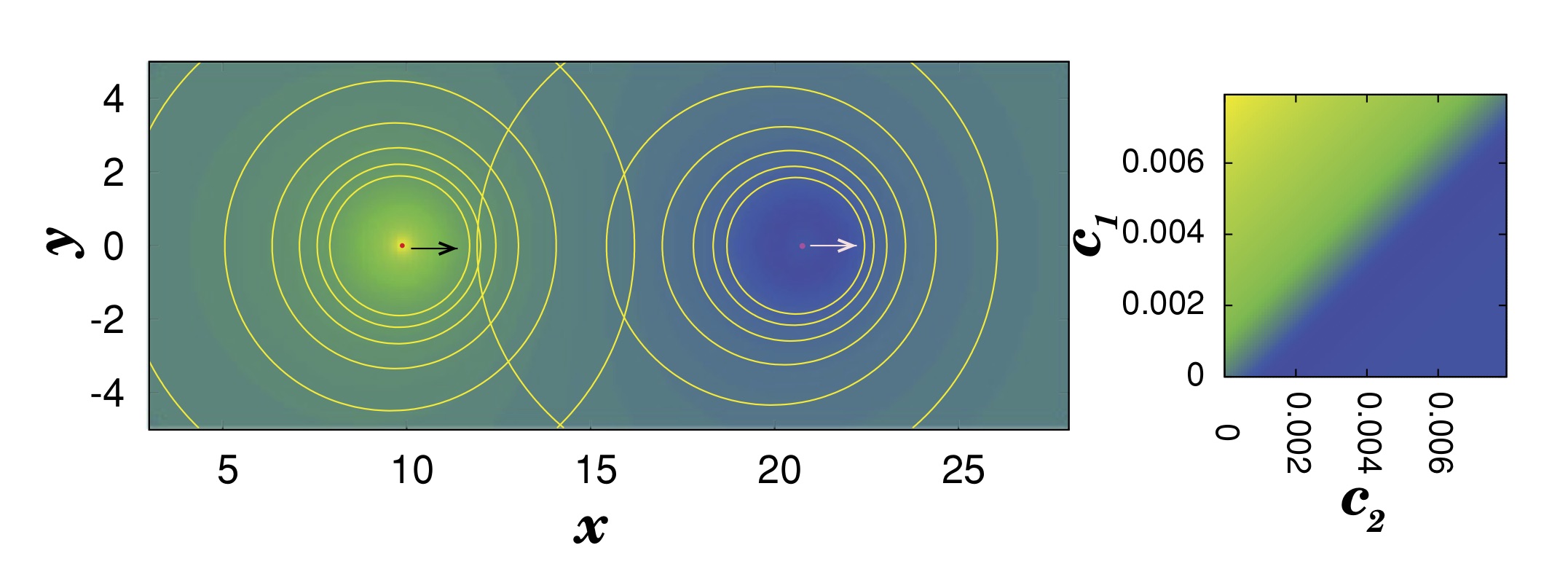}
  \caption{(Left) A predator (red dot on left) chases a prey (red not visible dot on right), while the latter tries to 
escape through chemotactic gradient sensing of the diffusing chemicals. The arrows indicate their respective direction 
of motion in the absence of fluctuations. The contours around each microbe represent the equiconcentration lines 
of the secreted chemicals in a two-dimensional projected plane in this case, indicating the asymmetry of the distribution. 
The color code used here for the spatial distribution of the secreted chemorepellant ($c_1$) and the chemoattractant ($c_2$), 
as they mingle in space, is shown in the right panel. Figure from Ref.\cite{sengupta2011chemotactic}.}
  \label{Fig5}
\end{figure}

A typical snapshot in the noise-free case is shown in Figure \ref{Fig5} which highlights the two concentration
fields and the predator (left particle) following the prey (right particle).

%Figure 5: hier Figure 1 von \cite{Sengupta2} einfügen mit Caption plus:
%From Ref.\ \cite{Sengupta2}.
As a result of the analysis performed in Ref. \cite{sengupta2011chemotactic}, there are basically two dimensionless
parameters which govern the escape and chase scenario, still in the case of vanishing noise. The first parameter is the
reduced length scale $\Delta*$ depending on $r_{12}$ 
which is the steady-state distance between the particles; the second parameter is 
a sensibility ratio $\delta$; see \cite{sengupta2011chemotactic} for details. The state diagram is shown in Figure 6
in the parameter plane, spanned by $\Delta^*$ and $\delta$, and shows regions of escape and capture.
For $\delta >1$ there is always trapping, independent of the initial conditions.
For $\delta <1$ it depends on the initial particle separation $r_0^*$ (see Figure 6b) which
effective $\Delta^*$ is realized. On the separation line there is steady state motion with a constant
particle separation, i.e. both particles move with the same speed. This line 
can be calculated analytically and is given by

\begin{equation} \label{eq:IV4}
\delta(\Delta^*)=(1+\Delta^{*-1}) \exp{(-\Delta^{*-1})}
\end{equation} 
%Hier Gleichung (7) on Ref. \cite{Sengupta2} einfügen (IV.4)

\begin{figure}[]
  \centering
     \includegraphics[width=1.0\textwidth]{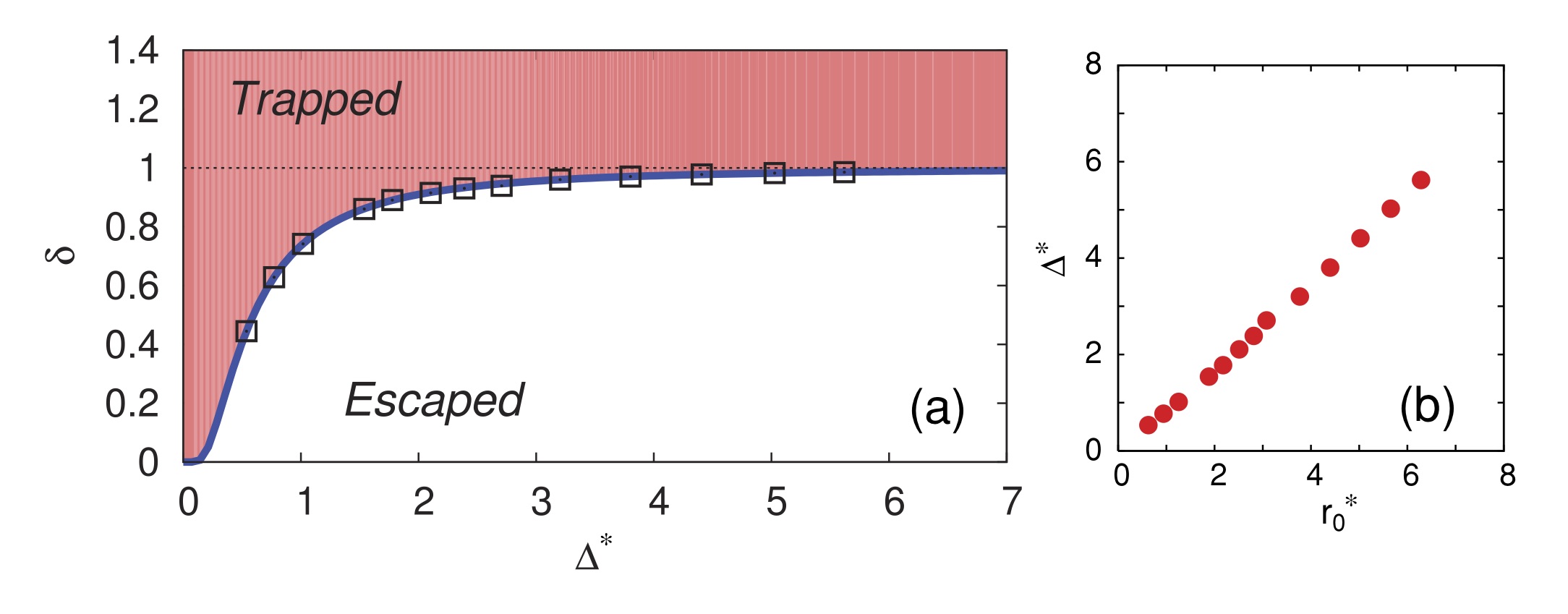}
  \caption{(a) Dynamical phase diagram of the chemotactic predator-prey system, constructed in the $\Delta^*-\delta$ parameter space, showing the trapped (shaded) and escaped phases. The phase boundary (thick solid line) is obtained analytically and matches the simulation data (boxes). The horizontal thin dotted line ($\delta=1$) represents the upper bound for the trapped-to-escaped dynamical phase transition (see text). (b) The dependence of the catching range ($\Delta^*$) on the initial separation ($r_0^*$), as obtained from simulations. 
Figure from Ref.\cite{sengupta2011chemotactic}.}
  \label{Fig6}
\end{figure}
%Figure 6: hier Figure 2 von \cite{Sengupta2} einfügen mit Caption plus:
%From Ref.\ \cite{Sengupta2}.

The particular form of the diffusing chemical field in the absence of evaporation, Eq.~(\ref{eq:II13}),
allows the prediction of two scaling laws. The first one applies to escape situations and shows that 
the distance $x$ between the two particles
increases subdiffusively with time $t$ as $t^{1/3}$, while the second one applies to trapping situations and predicts that the distance between predator and prey decreases 
as $(t_{trap} -t)^{1/3}$
with a finite trapping time $t_{trap}$ where $x=0$.

As a final remark, there are also other predator-prey models which model the predator and prey dynamics
either on a lattice
\cite{Oshanin} or with
more complex models designed for real bacteria \cite{Presse}. A marvellous realization
of a synthetic predator-prey system employs a pair of an ion-exchange resin and a passive 
charged colloidal particle which moves autonomously as a "modular swimmer" \cite{Palberg_recent};
another interesting realization of a predator-prey-like pair is based on two different particles, which are actuated as a pair \cite{Liebchen}.

\section{Collective behavior of few and many active particles} %V

Tsori and de Gennes \cite{Tsori} have pointed out the analogy between 
chemoattractive matter and  {\it gravity}
in three spatial dimensions. In fact, the chemotactic coupling (\ref{eq:II14}) together with the Coulomb orbitals (\ref{eq:II8}) and the
linearity of the diffusion equations implies that a one-component system is identical to gravitating particles.
The dynamics will lead to clustering and finally to a collapse ("black hole").
Hence one can study aspects of the dynamics of a black hole collapse scaled down in a Petri dish, 
see e.g. Ref. \cite{Oettel_Petri} for a similar idea.

{\it Binary mixtures\/} of particles with chemotactic coupling coefficients of opposite sign 
lead to a similar physical behavior as 
oppositely
electrically charged mixtures. These systems form interesting cluster structures and lead to effectively 
non-reciprocal forces, i.e. they break Newton's third law actio=reactio.
These nonreciprocal forces can lead to self-propulsion and self-rotation\cite{Bartnick_JPCM} which only emerges if different colloids closely approach each other and form 
"active molecules" appearing in a broad variety of shapes as movers, rotators and circle swimmers
\cite{Soto_Golestanian1,Soto_Golestanian2, Liebchen}. 
The simplest example of such an active molecule is a moving dimer similar to the predator-prey system discussed in the previous chapter.
A direct experimental confirmation of active molecules consisting of two species
was found in Ref. \cite{Palberg_Speck} where two different types of ion exchange 
resins provide the active constituents.

\begin{figure}[H]
  \centering
     \includegraphics[width=0.99\textwidth]{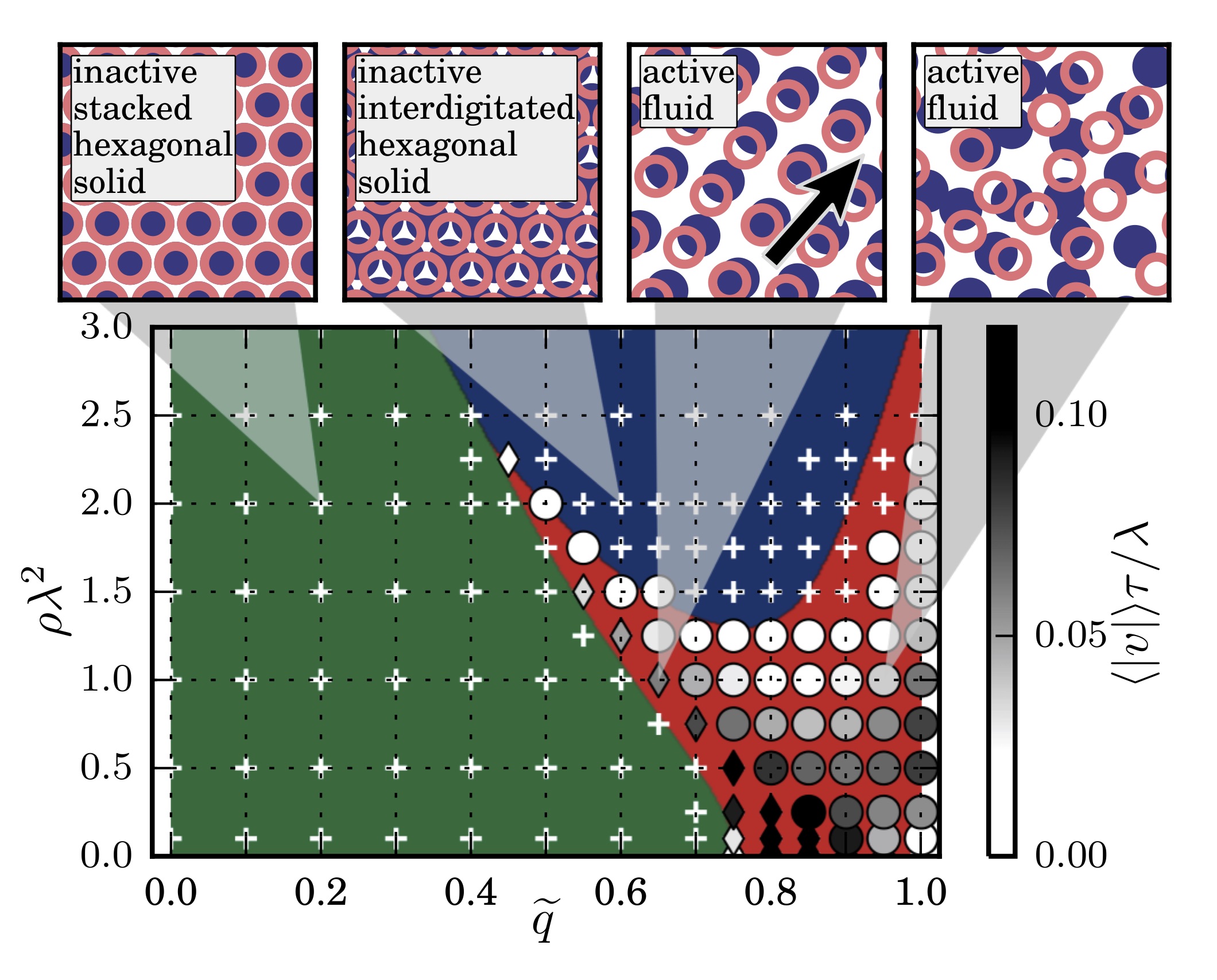}
  \caption{State diagram in the zero-temperature limit, plotted in the plane of a reduced number density $\rho$ and relative wake charge $\tilde{q}$ . Color coding depicts results obtained from the stability analysis, symbols show numerical results. Inactive systems (+) can be either \textit{stacked hexagonal solid} (green background) or \textit{interdigitated hexagonal solid} (blue background). For \textit{active fluid} regimes ($\circ$, red background), the average particle velocities are indicated by a gray scale. Diamonds ($\diamond$) are used instead of circles if active doublets emerge whose decay time $\tau_{D}$ exceeds a threshold of $10^{3} \tau$. The states are illustrated by typical snapshots. Figure from Ref.\cite{Bartnick_JCP}.}
  \label{Fig7}
\end{figure}

The full many-body behavior of a binary mixture interacting with chemotactic-based 
nonreciprocal interactions was studied
in Refs. \cite{Bartnick_JCP}, however in the different context of complex plasmas. 
In Ref. \cite{Bartnick_JPCM}, the connection
to chemotaxis was worked out explicitly.
One example for the collective behaviour in a binary system of chemotactically coupled species is
shown in Figure \ref{Fig7} at vanishing noise. On the $x$-axis a relative wake charge is plotted which 
corresponds to the non-reciprocity governed by the asymmetry in the sensing mechanisms of the two particles.
The $y$-axis shows the two-dimensional particle density. There is a rich steady state diagram
 with four different dynamical states involving inactive (i.e. non-moving) states and active ones.
The latter are either swarms or orientationally disordered active fluids.

%Figure 7: hier Figure 5 of \cite{Bartnick_JCP} einfuegen mit Caption

A further interesting setup of particles interacting via chemotaxis is provided by autophoretic Janus colloids. 
These particles catalyze a chemical reaction on part of their surface only resulting in a 
chemical gradient across their own surface. This self-produced gradient sets them into motion via diffusiophoresis or a similar mechanism. 
Remarkably, a Janus colloid does not only respond to self-produced gradients, but also to
gradients produced by other Janus colloids, essentially by chemotaxis (or taxis with respect to another
phoretic field). Thus, phoretic Janus colloids interact ``chemically'' and provide a synthetic analogon to microbiological signaling \cite{Saha2014,Lutz,Liebchen2015}. 
These chemical interactions play a crucial role for the collective behaviour of large suspensions of Janus colloids \cite{Liebchen2017} and
can generate patterns including clusters \cite{Saha2014,Pohl2014,Pohl2015,Liebchen2015,Liebchen2017}, traveling waves \cite{Liebchen2017} and continuously moving 
patterns \cite{Liebchen2015} and in case of chiral active particles also spiral patterns  
and phase separation with traveling waves emerging within the dense phase \cite{Liebchen2016}.

\section{Conclusions} %VI

In conclusion, we have discussed models for chemotactic behavior of microorganisms and synthetic particles 
in a diffusing chemical concentration field focusing on three different scenarios:
i) autochemotaxis of a single particles, ii) predator-prey models arising from chemotactic
coupling to two different chemicals secreted by the predator and the prey, iii) clusters and collective behavior of
many particles coupled via their chemotactic response to chemical fields produced by other particles. Here, 
attractive autochemotaxis may lead to self-localization while
repulsive autochemotaxis leads to trajectories avoiding their own past.
%gives rise to ballistic self-propulsion, at least for intermediate time scales.
We have also discussed that a moving chemotactic source leads to a front-rear asymmetry in the chemical field resulting in 
marked scaling laws for predator-prey systems.
Finally, chemotaxis in multi-species systems provides an avenue towards 
a new world of active molecules where we have just started to tap the full potential
of the novel cluster formation processes. Collective behavior includes a swarming of chemotactically
coupled particles at finite concentrations.

We close with an outlook to future problems. First, at high chemical concentration 
or strongly coupled chemical fields, the simple diffusion equation picture will break down, calling for new models. 
This is in particular important for multivalent
microions at high concentrations. Recent developments have considered these effects of 
strong coupling in using nonlinear diffusion equations.
These equations can be based on the dynamical version of classical density functional theory, so-called dynamical
density functional theory (DDFT) \cite{Marconi,Archer_Evans,Espanol_Loewen,my_DDFT_review}.
In this framework, the equations of motion of a chemical around a point source are given by

\begin{equation} \label{eq:V1}
\frac{\partial c(\vec{r}, t)}{\partial t} = D \vec{\nabla} c(\vec{r},t) \vec{\nabla} \frac{\delta \mathcal{F} \left[ c(\vec{r},t)\right] }{\delta c(\vec{r},t)} - \mu c(\vec{r},t) + \lambda_e \delta(\vec{r})
\end{equation} 

where ${\cal{F}} [n([\vec{r}]$ is the equilibrium free energy density density functional \cite{Evans,Singh,Loewen94}.
For a noninteracting system (ideal gas), the functional is known explicitly and in this limit
 the traditional diffusion equation (\ref{eq:II1}) is recovered. Nontrivial particle correlations as arising
from interactions among the chemical species are contained in the functional in the general case and 
make the diffusion equation nonlinear.
In certain cases, linearization is possible and corresponding analytical solutions
for the Green's function of diffusing interacting particles can indeed be found within DDFT \cite{Heinen_EPJST}.

A second important generalization concern time-dependent secreting 
rates as embodied in a non-constant function  $\lambda_e(t)$
such as e.g. an emission rate that is periodic in time. This situation has recently been considered \cite{Heinen_EPJST}
and leads to propagating density waves of the chemical around the emitting source. Again, in some special cases, the
Green's function can be found analytically within DDFT \cite{Heinen_EPJST}.

Third, in terms of predator-prey models, the situation of a single predator and a single prey can be
generalized towards a herd of prey and to a group of chasers. This has been
discussed in the literature within different models, see e.g. \cite{single_predator_herd_of_prey,group_chase0,group_chase1,group_chase2}
but needs to be extended within the chemotactic context. Efficient chase and escape 
strategies \cite{Benichou_RMP} may depend on the details of
predator/prey perception.

Fourth, most of our consideration were done in the bulk. Confinement near system walls and crowding situations 
will change both the diffusion of the chemical as well as the chemotactic dynamics.
Similarly, chemotaxis in complex environments, such as traveling waves may lead to interesting transport effects \cite{Geiseler2016}.
We are just at the beginning of a systematic understanding 
of chemotaxis in complex environment. 

Finally we have considered the evaporation of different chemicals by a constant rate
 in our modelling. If two different chemicals
which e.g. govern a predator-prey system or the formation of the active molecule, 
react between themselves this would constitute
a more complicated and highly interesting problem with new scenarios induced by coupling 
nonlinear reaction-diffusion equations
to the chemical kinetics and the particle motions.

\section{Acknowledgements}
H.L. gratefully acknowledges support by the Deutsche
Forschungsgemeinschaft (DFG) through grant
LO 418/19-1.

\bibliographystyle{ws-rv-van}
\bibliography{ws-rv-sample}

%\blankpage
%\printindex[aindx]                 % to print author index
%\printindex                         % to print subject index

\end{document}